\documentstyle[12pt,epsf]{article}
\newcommand{\noi}{\noindent}
\newcommand{\eq}{\begin{equation}}
\newcommand{\en}{\end{equation}}
\newcommand{\eqa}{\begin{eqnarray}}
\newcommand{\ena}{\end{eqnarray}}

\newcommand{\tr}{\mbox{Tr}\,}

\newcommand{\caa}{ {\cal A}}

\newcommand{\cak}{ {\cal K}}

\newcommand{\bpartial}{{\bar \partial}}

\newcommand{\ageq}{\mbox{}_{\textstyle \sim}^{\textstyle > }}

\newcommand{\lra}{\longrightarrow}

\begin{document}

\renewcommand{\theequation}{\arabic{section}.\arabic{equation}}
\renewcommand{\thesection}{\arabic{section}}
\renewcommand{\thefootnote}{\fnsymbol{footnote}}

\vspace{-2cm}

\hbox{}
\noindent   April 2002        \hfill

\vspace*{1.0cm}

\begin{center}

{\Large Gluon propagators and the choice of the gauge field}

\vspace{3mm}
{\Large in $SU(2)$ theory on the lattice}
\footnote{
Work supported by the grant INTAS-00-00111 and RFBR grant 02-02-17308.
}

\vspace*{0.8cm}

I.L. Bogolubsky$\mbox{}^a$ and V.K.~Mitrjushkin$\mbox{}^{a,b}$

\vspace*{0.3cm}

\end{center}

{\sl \noindent
 \hspace*{6mm} $^a$ Joint Institute for Nuclear Research, 141980 Dubna,
          Russia \\
 \hspace*{6mm} $^b$ Institute of Theoretical and Experimental Physics,
                    Moscow, Russia}

\vspace{0.5cm}
\begin{center}
{\bf Abstract}
\end{center}

We study numerically magnetic $G_M(p)$ and electric $G_E(p)$ gluon
propagators and their dependence on the choice of the lattice gauge
field $\caa_{x\mu}$ in $SU(2)$ gauge theory, especially, in the
low--momentum limit. We find that two different $\caa_{x\mu}$
definitions are equivalent up to a trivial renormalization of the
propagator, at least, in the main approximation.

\section{Introduction}

Gauge variant Green functions are supposed to contain an important
information about the large distance physics and mechanism(s) of
confinement. For example, at zero temperature the expected infrared
suppression of the gluon propagator in the Lorentz (or Landau) gauge is
believed to be connected to the Gribov's confinement scenario
\cite{grib,zwa2} (see also \cite{dses}). In the high temperature
(chromoplasma) phase the large distance and low--momentum dependence of
the gluon propagator may provide valuable information about the electric
and magnetic screening mechanisms \cite{lind,gpya}.

On the lattice the definition of gauge variant Green functions, e.g.
gluon propagators, is not free from the ambiguity.
Indeed, the lattice gluon propagator $G_{\mu\nu}$ is given by

\eq
G_{\mu\nu}(x-y) \sim \left\langle \tr
\Bigl( \caa_{x\mu}\caa_{y\nu}\Bigr)\right\rangle~,
\en

\noi where $\caa_{x\mu}$ are gauge fields on the lattice and
$~\langle\ldots\rangle~$ means the statistical average with some gauge
fixing condition. There is no
unique definition of $\caa_{x\mu}$ on the lattice in terms of the link
variables $U_{x\mu}$. This problem has been addressed recently in
\cite{rom1,cume,cuka,nafy} and some possible choices of $\caa_{x\mu}$ have
been discussed.

A similar ambiguity exists in the choice of the gauge which is the
Lorentz gauge in our case. On the lattice there are different ways to
choose the Lorentz gauge fixing condition, all of them being the same
in the continuum limit.

In this paper we study the space--space (magnetic) $~G_M(p)~$ and
time--time (electric) $~G_E(p)~$ correlators in the high temperature
plasma phase with a special attention paid to their low--momentum
behaviour.  The main point in our study is the dependence of the
correlators on the choice (definition) of lattice gauge field
$\caa_{x\mu}$ as well as on the specific lattice implementation of
Lorentz gauge fixing condition.

In what follows the periodic boundary conditions are used, the lattice
size is $V_4 = N_4\times N_s^3$ and $~\bpartial_{\mu}~$ is a backward
lattice derivative. We use also $~\cak_{\mu}(p) = \frac{2}{a}
\sin\frac{ap_{\mu}}{2}~$  and $~\cak^2(p) =\sum_{\mu} \cak_{\mu}^2(p)~$,
$a$ being a lattice spacing.

\section{Gauge fields and Lorentz gauge conditions}
\setcounter{equation}{0}

\subsection{Definitions of the gauge fields}

The standard Wilson action \cite{wils} with $SU(2)$ gauge group is

\eq
S(U) = \beta \sum_x\sum_{\mu >\nu}
\left[ 1 -\frac{1}{2}\tr \Bigl(U_{x\mu}U_{x+\mu;\nu}
U_{x+\nu;\mu}^{\dagger}U_{x\nu}^{\dagger} \Bigr)\right]~;
\qquad \beta = \frac{4}{g^2}~,
\en

\noi where $g$ is a bare coupling constant and $U_{x\mu}\in SU(2)$ are
link variables. Under gauge transformations $ \Omega $ field variables
$U_{x\mu}$ transform as follows

\eqa
U_{x\mu} &\stackrel{\Omega}{\lra}& U_{x\mu}^{\Omega}
= \Omega_x U_{x\mu}\Omega_{x+\mu}^{\dagger}~;
\qquad \Omega_x \in SU(2)~.
           \label{gaugetr}
\ena

\noi A standard  definition \cite{mog1} of the lattice gauge field
$\caa_{x\mu}$ in terms of link variables $U_{x\mu}$ is

\eq
\caa^{(1)}_{x\mu} = \frac{1}{2iag}\Bigl( U_{x\mu}-U_{x\mu}^{\dagger}\Bigr)~,
              \label{field1}
\en

\noi However, this definition is not unique.  For example,  instead of
field $\caa^{(1)}_{x\mu}$  one can define gauge field
$\caa_{x\mu}^{\prime}~$ \cite{rom1}

\eq
\caa_{x\mu}^{\prime} =
\frac{1}{4iag}\Bigl( U_{x\mu}^2-(U_{x\mu}^{\dagger})^2\Bigr),
\en

\noi or another gauge field $\caa_{x\mu}^{\prime\prime},~$ \cite{cume}

\eq
\caa_{x\mu}^{\prime\prime} = \frac{4}{3}\caa_{x\mu}^{(1)}
-  \frac{1}{3}\caa_{x\mu}^{\prime}~,
\en

\noi etc. On the other hand, the proof of the (naive) continuum limit
existence of the lattice action $S(U)$ as well as the construction of
the lattice perturbation theory is based on the representation of the
link variable $U_{x\mu}$ in the form

\eq
U_{x\mu} \equiv \exp\Bigl\{ iag A_{x\mu}\Bigr\}~,
\qquad |{\vec \caa_{x\mu}}| \le\frac{2\pi}{ag},
              \label{field2}
\en

\noi and the expansion in series in powers of coupling constant $g$.
Therefore, from the point of view of the analytical (e.g., perturbative)
studies, there is a natural definition of the lattice gauge field,
$\caa_{x\mu}$

\eq
\caa^{(2)}_{x\mu} = A_{x\mu}(U_{x\mu})~,
\en

\noi $A_{x\mu}$ being defined in eq.(\ref{field2}). Given any group
element $~U = c_0 {\hat 1} + i{\vec \sigma}{\vec c} = \exp\{
\frac{i}{2}{\vec\theta}{\vec\sigma} \}~$, one can easily find that

\eq
{\vec\theta} = \frac{{2\vec c}}{|{\vec c}|}\arccos c_0~,
\en

\noi unless $~c_0\ne -1$ (the case $~c_0=-1~$ has measure zero and can
be discarded). A formal expansion in powers of spacing $a$ gives

\eq
\caa^{(1)}_{x\mu} = \caa^{(2)}_{x\mu} + O(a^2)~,
\en

\noi so that both definitions are equivalent in the naive continuum
limit.

In the rest of this paper the lattice spacing $a$ is chosen to be unity.

\subsection{Choice of the Lorentz gauge fixing condition}

In lattice calculations the usual choice of the Lorentz gauge condition
is \cite{mog1}

\eq
\sum_{\mu=1}^4 \bpartial_{\mu} \caa_{x\mu}^{(1)} = 0,
                         \label{gf1}
\en

\noi which is equivalent to finding an extremum of the functional
$F_U^{(1)}(\Omega)$,

\eq
F_U^{(1)}(\Omega) = \frac{1}{4V_4} \sum_{x\mu} \frac{1}{2}
\tr U^{\Omega}_{x\mu}~,
                  \label{gf2}
\en

\noi with respect to gauge transformatins $\Omega_x~$.  In what follows
this gauge is referred to as $LG_0$. Equally, one can choose another
form of the Lorentz gauge fixing condition (see also \cite{nafy}) which
will be referred to as $LG_1$ :

\eq
\sum_{\mu=1}^4 \bpartial_{\mu} \caa^{(2)}_{x\mu} = 0.
                         \label{gf3}
\en

\noi Evidently, both $LG_0$ and $LG_1$ are the same in the continuum limit.
The $LG_1$ gauge condition is equivalent to finding an extremum of the
functional $F_U^{(2)}(\Omega)$

\eq
F_U^{(2)}(\Omega) = \frac{1}{4V_4} \sum_{x\mu} \frac{1}{2}
\tr (A^{\Omega}_{x\mu})^2~.
                  \label{gf4}
\en

\noi The field $~A_{x\mu} = A_{x\mu}(U_{x\mu})$ shows the nonanalytic
dependence on the link variable $U_{x\mu}$. However, the functional
$F_U^{(2)}(\Omega)$ still remains a continuous function of $\Omega$.
Under the infinitesimal gauge transformation $~\delta\Omega_x =\exp\{
i\delta\omega_x\}~$  the variation of the functional $F_U^{(2)}(\Omega)$
is

\eq
\delta F_U^{(2)}(\delta\Omega) = - \sum_a \varphi_x^a \cdot
\delta\omega_x^a + \ldots~,
             \label{gcond_a}
\en

\noi where

\eq
\varphi_x^a = \sum_{\mu} \bpartial_{\mu} A_{x\mu}^a(U_{x\mu})~.
             \label{gcond_b}
\en

\noi Eqs.
(\ref{gcond_a}),(\ref{gcond_b}) define the numerical algorithm of gauge
fixing. To maximize $~F^{(2)}_U(\Omega)~$  one can choose the gauge
transformation matrix $\Omega_x=\exp\{i\omega_x\},$

\eq
\omega^a_x = -b\cdot \varphi^a_x~;
\qquad b>0~,
           \label{alg2}
\en

\noi successfully at all lattice sites $x$.  After a certain number of
gauge fixing sweeps a local maximum $~F^{(2)}_{max}(U)~$ of the
functional $~F^{(2)}_U(\Omega)~$ is reached with given accuracy.  The
value of the parameter $b$ in Eq.(\ref{alg2}) should be tuned to
optimize the convergence.

\subsection{Lorentz gauge and Gribov copies}

It is well--known that gauge fixing on the lattice as well as in the
continuum is affected by the existence of Gribov copies \cite{grib}.  If
one repeatedly subjects a configuration $~\bigl\{U_{x\mu}\bigr\}~$ to a
random gauge transformation as in Eq.(\ref{gaugetr}) and subsequently
applies to it the Lorentz gauge fixing procedure, one usually obtains
Gribov (or gauge) copies with  different values of $~F_{max}(U)$.  It is
a long--standing believe \cite{zwa1} that the ``true" gauge copy
corresponds to the absolute maximum of $~F_U(\Omega)$.

In our case the question of interest is the dependence of lattice
quantities in the Lorentz gauge(s) on the choice of the gauge copy. This
dependence has been well studied in a number of papers (see, e.g.
\cite{rom2,rom3,hkr1,cucc,larg}). It has been found that for different
choices of gauge copies variation of the gluon propagator is small and
comparable with statistical error.

Our study confirms this statement. For every equilibrium configuration
$~\bigl\{U_{x\mu}\bigr\}~$ we performed $10$ random gauge
transformations \footnote{It is worth noting that in some of the above
mentioned papers many more gauge copies have been used.} and then
applied a gauge fixing procedure. For the calculation of the gluon
propagator we have used either the first gauge copy of the equilibrium
configuration $~\bigl\{U_{x\mu}\bigr\}~$ or the ``best" one, i.e.  the
copy with the maximal value of $~F_{max}(U)$. The difference between the
(averaged) propagators appeared to be very small and negligible in
comparison with statistical errors. Therefore, we conclude that there is
no in fact Gribov copy problem, at least, for the $SU(2)$ lattice gluon
propagator in the Lorentz gauge.

Note that this situation is quite different from what one observes in
the case of $U(1)$ lattice gauge theory. Indeed, the $4d$ photon $U(1)$
propagator in the Lorentz gauge exhibits an extremely strong  dependence
on gauge copies \cite{napl,dube}.

\section{Numerical results}\setcounter{equation}{0}

We calculated magnetic $G_M(p)$ and electric $G_E(p)$ correlators,

\eq
G_M(p) = \frac{1}{2} \Bigl( G_{11}(p) + G_{22}(p)  \Bigr)~;
\qquad
G_E(p) = G_{44}(p)~,
\en

\noi where $G_{\mu\nu}(p)$ is given by

\eq
G_{\mu\nu}(p) =\frac{1}{2V_4} \left\langle \tr
\Bigl( \caa_{\mu}(p)\caa_{\nu}(-p)\Bigr)\right\rangle~.
               \label{prop1}
\en

\noi Field $\caa_{\mu}(p)$ is a Fourier transform

\eq
\caa_{\mu}(p) = \sum_x e^{-ipx-\frac{i}{2}p_{\mu}} \cdot \caa_{x\mu}~,
\en

\noi where $\caa_{x\mu}$ are defined in Eqs.~ (\ref{field1}) and
(\ref{field2}) with unit spacing.  Momenta $p$ have been chosen to be
directed along the third axis, i.e.  $~p=(0,0,p_3,0)~$ with $p_3 = 2\pi
n_3/N_s$ and $n_3 =0,~1,~\ldots~$.

Most of our calculations have been performed on $~4\times N_s^3~$
lattices with $~N_s= 16;~24;~32;~40~$ and on the $~N_4\times 24^3~$
lattices with $~N_4=6;~8~$.  The following values of $\beta$ have been
chosen: $\beta=2.512;~2.645$ and $2.74$. The choice of $\beta=2.512$
corresponds to the temperature $T=2T_c$ on the lattice with $N_4=4$ as
well as the choice of $\beta=2.645$ on the lattice with $N_4=6$ and
$\beta=2.74$ on the lattice with $N_4=8$ \cite{hkr2} \footnote{We are
grateful to F.  Karsch and J.  Engels for providing us the $\beta$-value
for $N_4=6$}.

As it has been already mentioned, at zero temperature one expects the
infrared suppression of the gluon correlator, i.e. the nonperturbatively
calculated gluon correlator is expected to be less singular than the
perturbative one. It has been shown that Lorentz (and Coulomb) gauge
fixed gluon correlators on the infinite lattice vanish in the zero
momentum limit due to the proximity of the Gribov horizon \cite{zwa2}.
(This has been confirmed recently in the framework of Dyson--Schwinger
equation formalism \cite{dses}).  Recently this study has been extended
to the finite temperature case \cite{zazw}. It has been shown that in
the infinite volume limit magnetic correlation functions vanish in the
infrared unlike the electric correlation functions \cite{zazw}.

\begin{figure}[pt]
\begin{center}
\leavevmode
\hbox{
\epsfysize=14cm
\epsfxsize=14cm
\epsfbox{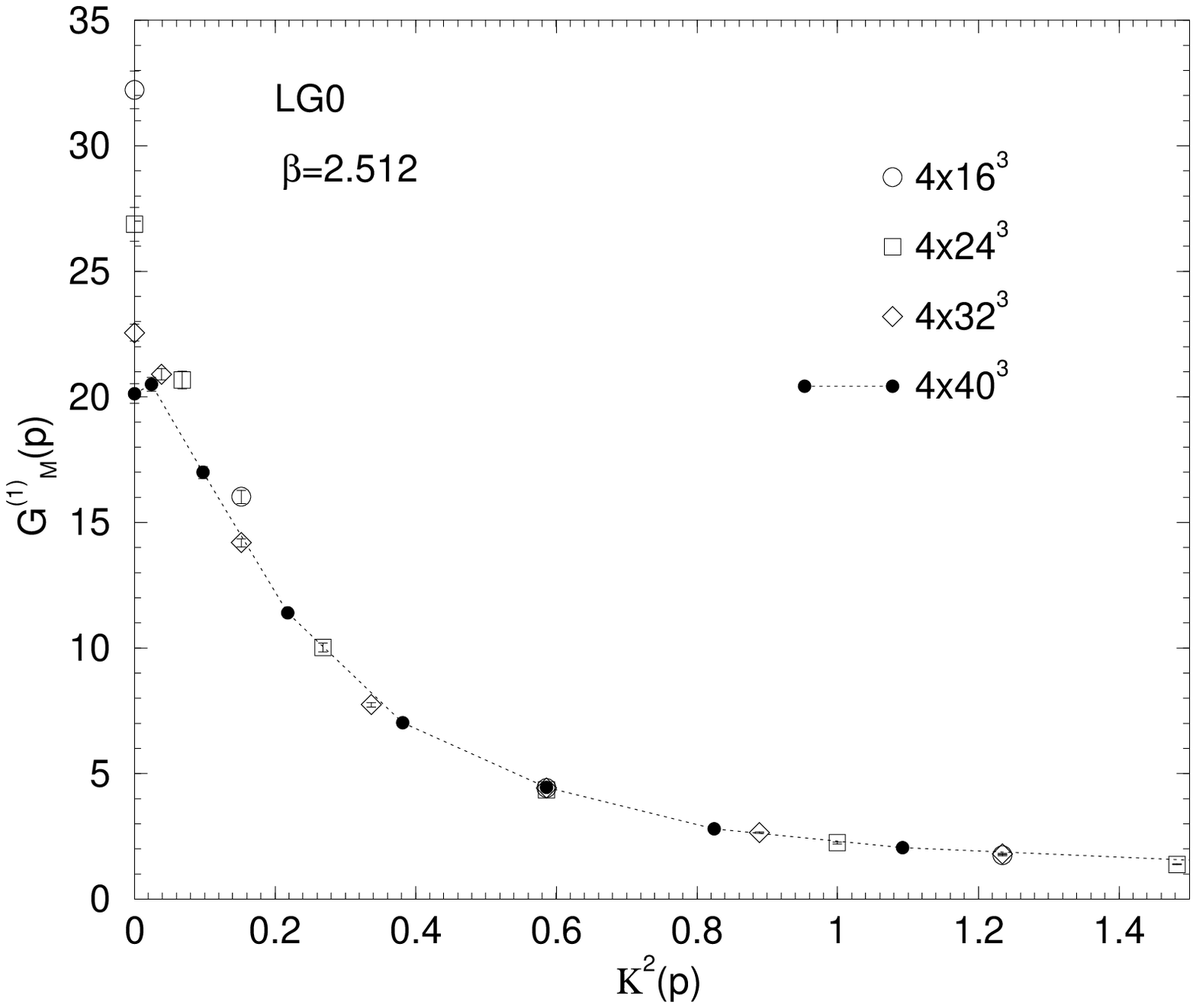}
}
\end{center}
\caption{$G^{(1)}_M(p)$ at $\beta=2.512$ for $LG_0$.
The line is to guide the eye.
}
\label{fig:cs_4latt_b2p512_LG0}
\end{figure}

\begin{figure}[pt]
\begin{center}
\leavevmode
\hbox{
\epsfysize=14cm
\epsfxsize=14cm
\epsfbox{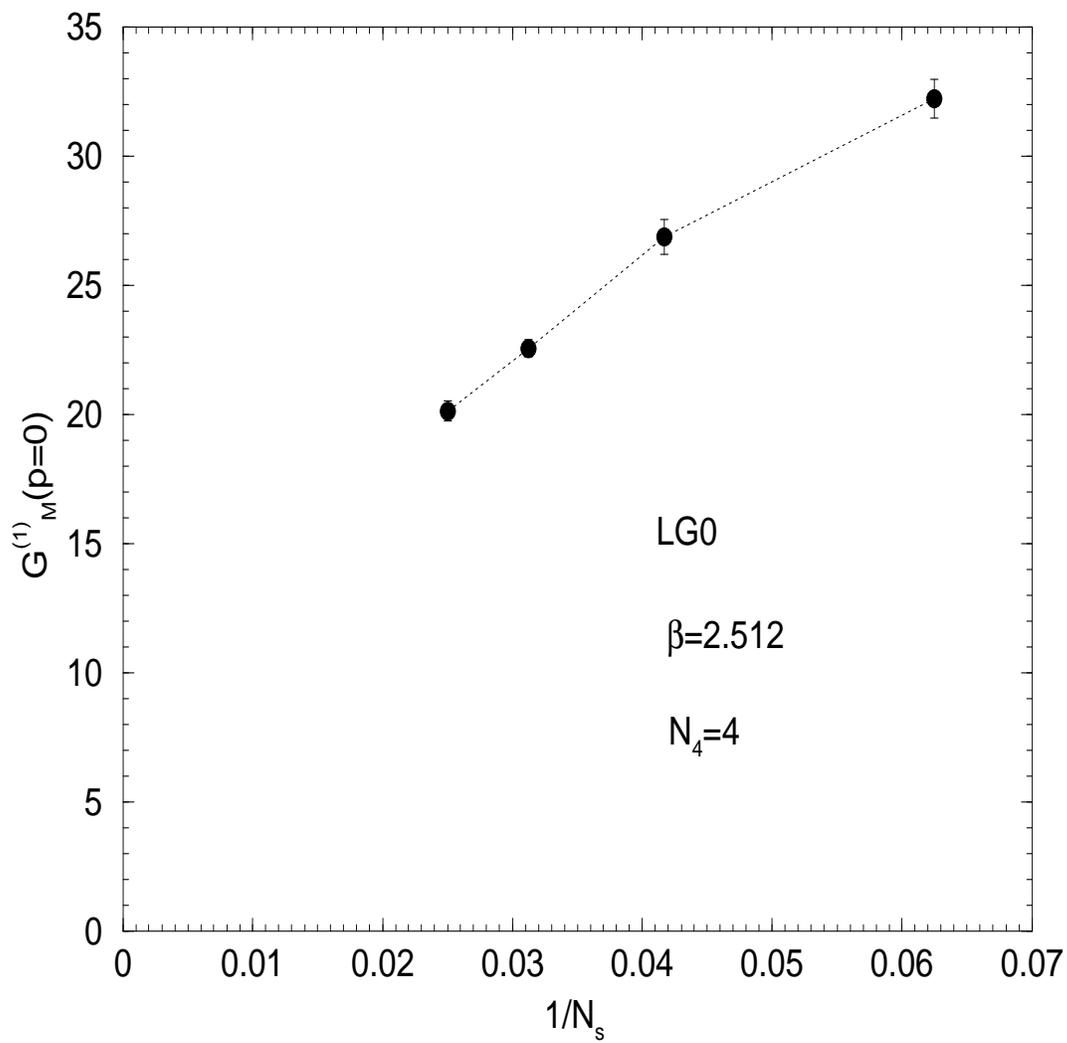}
}
\end{center}
\caption{The dependence of $G^{(1)}_M(0)$ on $N_s$ at $\beta=2.512$
for $LG_0$.
The line is to guide the eye.
}
\label{fig:zeromom_4latt_b2p512_LG0}
\end{figure}

In Figure \ref{fig:cs_4latt_b2p512_LG0} one can see the momentum
dependence of the magnetic correlator $G^{(1)}_M(p)$ for different
lattice sizes. At larger values of the momentum ($\cak^2 \ageq 0.2$) the
finite volume dependence is practically absent. On the other hand, at
values $\cak^2 \sim 0$ this dependence is rather strong. For our largest
lattice with $N_s=40$ the momentum dependence acquires a local maximum
at some nonzero value of $p$.  A similar structure in the high
temperature phase has been observed recently in \cite{cukp} (compare
with the low temperature case \cite{larg,aus1,aus2}). In Figure
\ref{fig:zeromom_4latt_b2p512_LG0} we show the dependence of the zero
momentum correlator $G^{(1)}_M(0)$ on $N_s$ at $N_4=4$ and
$\beta=2.512$. One can see a monotonic descrease of $G_M(0)$ with
increase of the lattice size $N_s$. However, one needs much larger
lattices to judge on the infinite volume limit.

\begin{figure}[pt]
\begin{center}
\leavevmode
\hbox{
\epsfysize=14cm
\epsfxsize=14cm
\epsfbox{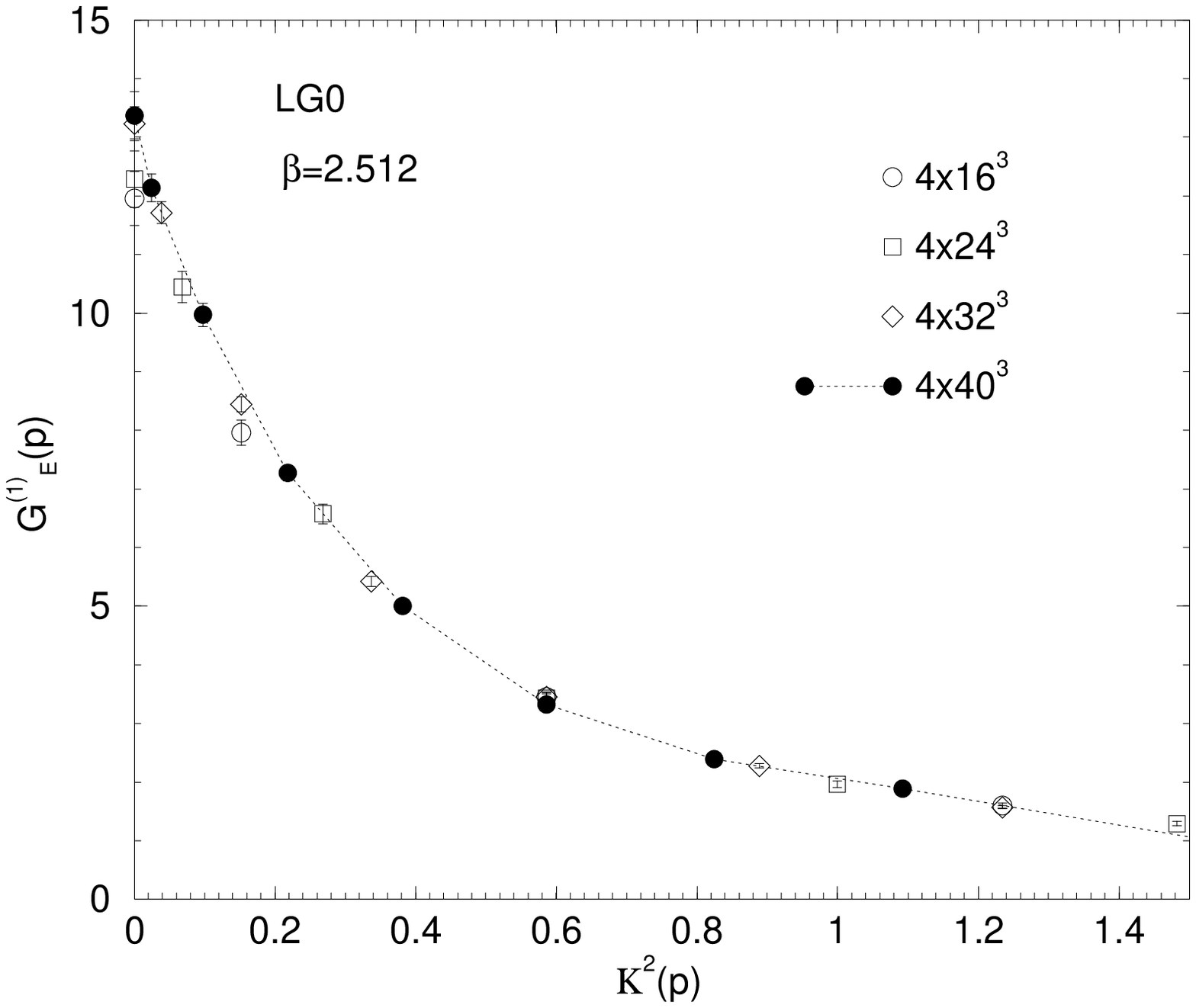}
}
\end{center}
\caption{$G^{(1)}_E(p)$ at $\beta=2.512$ for $LG_0$.
The line is to guide the eye.
}
\label{fig:c4_4latt_b2p512_LG0}
\end{figure}

The momentum dependence of the electric correlator $G^{(1)}_E(p)$ shown
in Figure \ref{fig:c4_4latt_b2p512_LG0} differs strongly from that of
the magnetic correlator, in agreement with \cite{zazw}. All curves are
monotonic and there is no any local maximum at nonzero momentum $~p~$.
Note also that at small momenta ($\cak^2 \sim 0$) the finite size
dependence of $G^{(1)}_E(p)$ is rather weak, in contrast to
$G^{(1)}_M(p)$ case.

To see the dependence of the correlators on the choice of the lattice
gauge fields we defined ratios $R_M$ and $R_E$ :

\eq
R_M(p;g^2;T) = \frac{G^{(1)}_M(p)}{G^{(2)}_M(p)}~;
\qquad
R_E(p;g^2;T) = \frac{G^{(1)}_E(p)}{G^{(2)}_E(p)}~.
\en

\begin{figure}[pt]
\begin{center}
\leavevmode
\hbox{
\epsfysize=14cm
\epsfxsize=14cm
\epsfbox{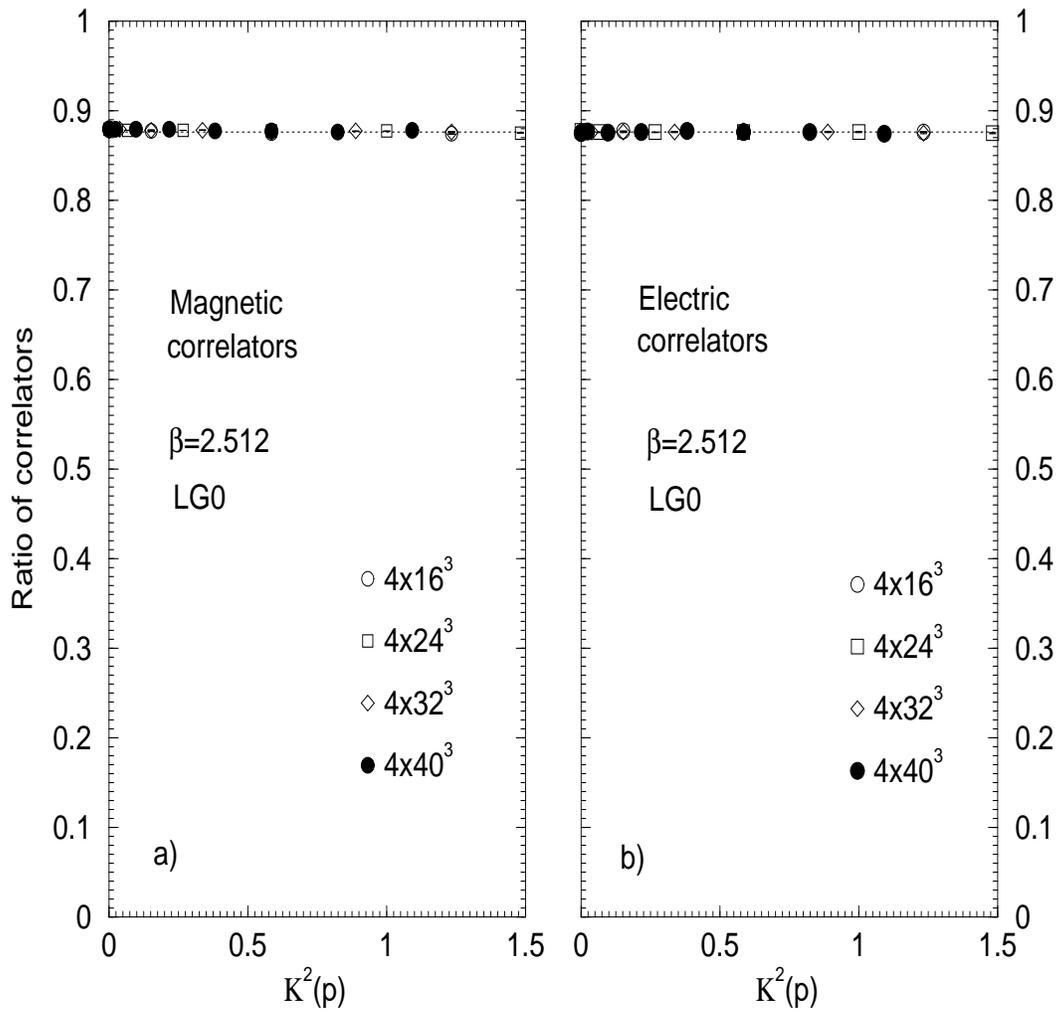}
}
\end{center}
\caption{Ratios of magnetic ({\bf a}) and electric ({\bf b})
correlators at $\beta=2.512$ for $LG_0$.
Lines are to guide the eye.
}
\label{fig:rc_4latt_b2p512_LG0}
\end{figure}

\noi These ratios are shown in Figures \ref{fig:rc_4latt_b2p512_LG0}a,b
for different lattices at $\beta=2.512$ , $LG_0$ and $LG_1$.  Up to
``second order" corrections (not visible in these Figures) both ratios
are  equal and momentum independent:

\eq
R_M(p;g^2;T) \simeq R_E(p;g^2;T) \simeq C(g^2;T)~.
\en

\begin{figure}[pt]
\begin{center}
\leavevmode
\hbox{
\epsfysize=14cm
\epsfxsize=14cm
\epsfbox{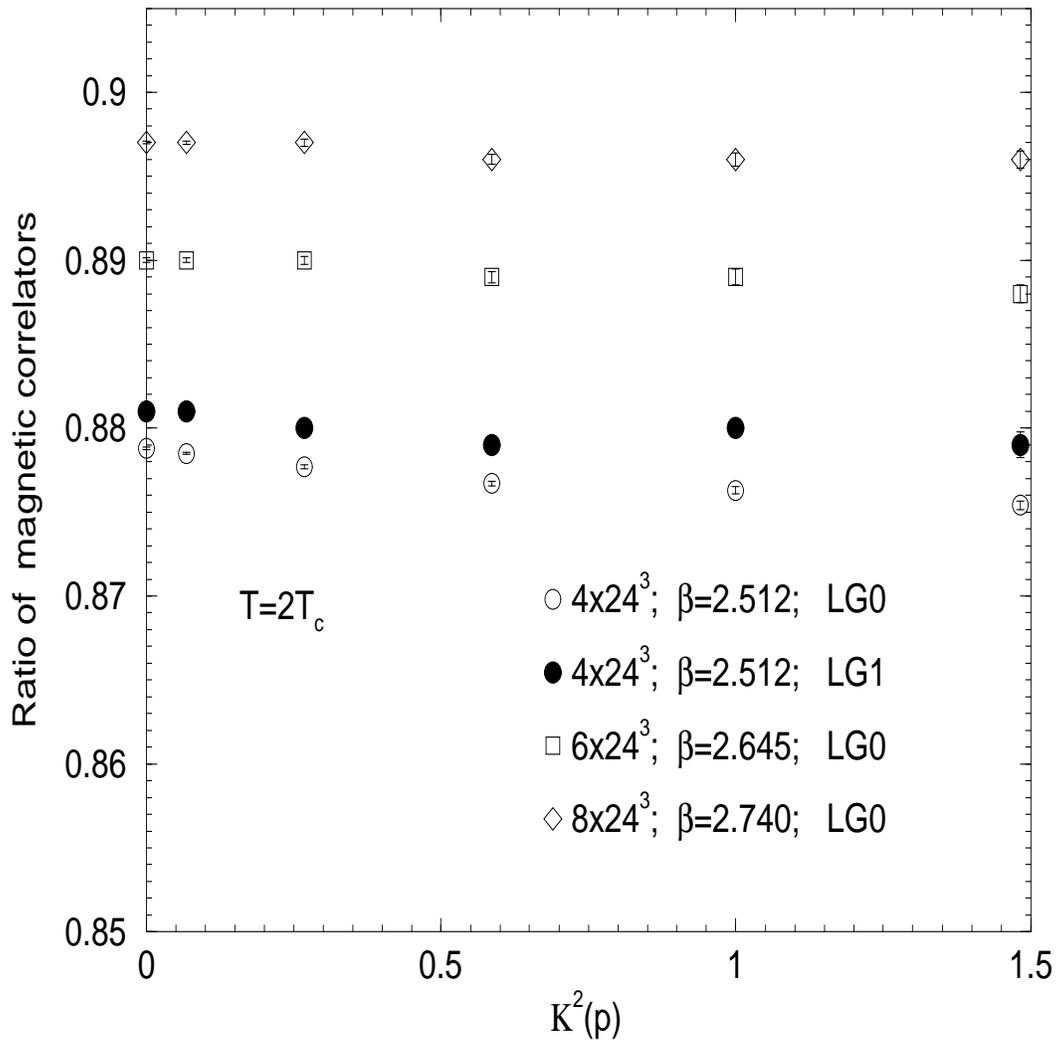}
}
\end{center}
\caption{Ratios of magnetic correlators at $\beta=2.512$
}
\label{fig:rc_3beta_LG0_LG1.eps}
\end{figure}

\noi A more detailed structure is shown in Figure
\ref{fig:rc_3beta_LG0_LG1.eps} for the ratio of magnetic correlators
under $LG_0$ and $LG_1$ at fixed temperature $T=2T_c$ and different
$\beta$'s.  One can see rather weak ($\sim 1\%$) dependence on $\beta$
and even weaker dependence on the choice of the gauge condition (i.e.
$LG_1$ vs. $LG_0$).  Another observation which can be made in Figure
\ref{fig:rc_3beta_LG0_LG1.eps} is a -- very weak -- momentum dependence
of the ratios. At the moment, we cannot exclude that this dependence is
a numerical artifact.  This point needs more detailed study.

\section{Conclusions} \setcounter{equation}{0}

We studied numerically the transverse magnetic $G_M(p)$ and electric
$G_E(p)$ gluon propagators in the high temperature plasma phase ($T >
T_c$) in the $SU(2)$ lattice gauge theory.

The transverse magnetic propagator $G_M(p)$ exhibits very strong volume
dependence in the infrared region. The value $G_M(0)$ tends to decrease
with volume increase in agreement with the prediction by Zahed and
Zwanziger \cite{zazw}. It is worthwhile to note that one needs rather
large lattices to study this effect in details.

The momentum dependence of the electric correlator is very much
different from that for the magnetic correlator. There is no any local
maximum at nonzero momentum $~p~$ and in the infrared region the finite
size dependence of $G_E(p)$ is rather weak, in contrast to $G_M(p)$
case.

We studied the dependence of the propagators on the choice of the
lattice gauge field $\caa_{x\mu}$, one of them being standard and
another one being ``natural" from the viewpoint of the perturbation
theory, as well as their dependence on the choice of the lattice Lorentz
gauge. We found that two different $\caa_{x\mu}$ definitions are
equivalent up to a trivial renormalization of the propagator, at least,
in the main approximation.  Most probably, this proportionality factor
can be explained (at least, partially) by the tadpole renormalization
(see also the discussion  in \cite{mtes}).


%
%

\vfill

%
%

\vfill

%
%

\vfill


%
%

\vfill


%
%

\vfill


\begin{thebibliography}{999}
\newcommand{\prd}[1]{Phys.~Rev.~{\bf D#1}\ }
\newcommand{\plb}[1]{Phys.~Lett.~{\bf #1B}\ }
\newcommand{\npb}[1]{Nucl.~Phys.~{\bf B#1}\ }
\newcommand{\prl}[1]{Phys.~Rev.~Lett.~{\bf #1}\ }
\newcommand{\prep}[1]{Phys.~Rep.~{\bf #1}\ }
\newcommand{\ap}[1]{Ann.~Phys.~{\bf #1}\ }
\newcommand{\cmp}[1]{Commun.~Math.~Phys.~{\bf #1}}
\newcommand{\rmp}[1]{Rev.~Mod.~Phys.~{\bf #1}}
\newcommand{\ptp}[1]{Prog.~Theor.~Phys.~{\bf #1}}
%
\bibitem{grib} V.N.~Gribov, \npb{139} (1978) 1.
\bibitem{zwa2} D. Zwanziger, Phys. Lett. {\bf B257} (1991) 168; Nucl. Phys.
{\bf B364} (1991) 127.
\bibitem{dses} L. von Smekal, R. Alkofer and A. Hauck, \prl{79} (1997)
3591; ~~\ap{267} (1998) 1.
\bibitem{lind} A.D. Linde, Phys. Lett. {\bf B96} (1980) 289.
\bibitem{gpya} D.J. Gross, R.D. Pisarski and L.G. Yaffe, \rmp{53} (1981)
43.
\bibitem{rom1} L. Giusti, M.L. Paciello, S. Petrarca, B. Taglienti
and M. Testa, Phys. Lett. {\bf B432} (1998) 196.
\bibitem{cume}  A. Cucchieri and T. Mendes, hep-lat/9902024.
\bibitem{cuka}  A. Cucchieri and F. Karsch,
Nucl.Phys.Proc.Suppl. {\bf 83} (2000) 357.
\bibitem{nafy} H. Nakajima and S. Furui, Nucl.Phys.Proc.Suppl.
{\bf 73} (1999) 865; \\
H. Nakajima, S. Furui and A. Yamaguchi, (2000) hep--lat/0007001.
\bibitem{wils} K. Wilson, \prd{10} (1974) 2445.
\bibitem{mog1}  J.E. Mandula and M. Ogilvie, Phys. Lett. {\bf B185} (1987) 127.
\bibitem{zwa1} D. Zwanziger, \npb{364} (1991) 127; \npb{378} (1992) 525.
\bibitem{rom2} M.L. Paciello, C. Parrinello, S. Petrarca, B. Taglienti
and A. Vladikas, Phys. Lett. {\bf B289} (1992) 405.
\bibitem{rom3} M.L. Paciello, S. Petrarca, B. Taglienti and A. Vladikas,
Phys. Lett. {\bf B341} (1994) 187.
\bibitem{hkr1} U.M. Heller, F. Karsch and J. Rank, Phys. Lett.
{\bf B355} (1995) 511.
\bibitem{cucc}  A. Cucchieri, \npb{507} (1997) 353.
\bibitem{larg}  K. Langfeld, H. Reinhardt and J. Gattnar,
\npb{621} (2002) 131.
\bibitem{napl} A. Nakamura and M. Plewnia, Phys. Lett. {\bf B255} (1991) 274.
\bibitem{dube}  V.G. Bornyakov, V.K. Mitrjushkin, M. Mueller--Preussker
and F. Pahl, Phys. Lett. {\bf B317} (1993) 596 ; \\
V.K. Mitrjushkin, Phys. Lett. {\bf B389} (1996) 713;  \\
I.L. Bogolubsky, V.K. Mitrjushkin, M. Mueller--Preussker and
P. Peter, Phys. Lett. {\bf B458} (1999) 102.
\bibitem{hkr2} U.M. Heller, F. Karsch and J. Rank, Phys. Rev. {\bf D57}
(1998) 1438.
\bibitem{zazw} I. Zahed and D. Zwanziger, \prd{61} (2000) 037501.
\bibitem{cukp}  A. Cucchieri, F. Karsch and P. Petreczky,
Phys. Lett. {\bf B497} (2001) 80;
\prd{64} (2001) 036001.
\bibitem{aus1}  D.B. Leinweber, J.I. Skullerud, C. Parrinello and A.G. Williams,
\prd{60} (1999) 094507; \prd{61} (2000) 079901.
\bibitem{aus2}  F.D.R. Bonnet, P.O. Bowman, D.B. Leinweber,
A.G. Williams and J.M. Zanotti,  Phys. Rev. {\bf D64} (2001) 034501.
\bibitem{mtes} M. Testa, JHEP {\bf 04} (1998) 002.


\end{thebibliography}
\end{document}